\documentstyle[12pt]{article} 
\begin{document}

\title{Hyperfine Effects\\and Large $1/{m}_{Q}^{2}$ Corrections}

\author{B. Holdom\thanks{holdom@gpu.utcs.utoronto.ca} \hspace{0.5em}
and \hspace{0.4em} M.
Sutherland\thanks{marks@medb.physics.utoronto.ca}
     \and
     {\small {\it Department of Physics, University of Toronto}} \\
     {\small {\it 60 St. George St., Toronto, Ontario, Canada M5S
1A7}}  } \date{UTPT-92-24; hep-ph/9212277} \maketitle

\begin{abstract} Effects associated with hyperfine mass splitting
between pseudo\-scalar  and vector mesons may induce surprisingly
large $1/{m}_{Q}^{2}$  corrections in the heavy quark expansion.  We
demonstrate this in a  relativistic quark model by calculating to all
orders in $1/{m}_{Q}$, with  and without hyperfine effects, and
comparing to the first order results.   Total corrections of 30\% or
more are quite possible in the decay rates  for $B\rightarrow Dl\nu$
and $B\rightarrow  D^{\ast}l\nu$ near zero recoil. \end{abstract}
\vspace{2ex}

The subject of semileptonic weak decays of mesons containing a heavy
quark has recently attracted much attention.  In the limit in which
the  heavy quark mass ${m}_{Q}$ becomes infinite the theoretical
treatment of  these decays is greatly simplified,\cite{1} potentially
reducing the  uncertainty in the extraction of the Kobayashi-Maskawa
elements.  But  more insight into the magnitude of the corrections to
this limit is  required.

We present here a relativistic quark model which provides a
representation of the decay amplitudes to all orders in $1/{m}_{Q}$.
In \cite{2} we show that the model correctly incorporates all heavy
quark symmetry relations at zeroth and first order in the $1/{m}_{Q}$
expansion.  We will use the model to study a source of  corrections
to the heavy quark limit which has not yet been seriously
considered.  This is the spin symmetry breaking effects due to gluon
exchange between the heavy and light quarks.  These hyperfine effects
split the vector and pseudoscalar masses, but they will also cause a
distortion of the light quark wave function in a way which depends on
the spin of the heavy quark.

Our relativistic model does not deal with gluon exchange directly,
but it relates the  distortions in light quark ``wave functions" to
the meson mass splitting.   In \cite{2} we explored the sensitivity
of amplitudes describing  $B\rightarrow Dl\nu$ and $B\rightarrow
D^{\ast}l\nu$ to this hyperfine  splitting at first order in the
heavy quark expansion.  Hyperfine splitting  had no effect on any of
the amplitudes at zero recoil.  This gives some  motivation for the
consideration of ${\cal O}(1/{m}_{Q}^{2})$ effects, in  addition to
the fact that the main source of corrections affecting the
extraction of ${V}_{cb}$ occurs at this order.

A systematic analysis of ${\cal O}(1/{m}_{Q}^{2})$ corrections in the
heavy quark effective theory has recently been made in \cite{6}.
Some  attempt was also made to estimate the quantities relevant to
semileptonic decay at zero recoil, and to do this the authors
appealed to  the ISGW nonrelativistic quark model.\cite{7}  But the
ISGW meson wave  functions are ``spin averaged"; hyperfine effects
are not included.  The  conclusion in \cite{6} is therefore
consistent with ours: if hyperfine  effects are ignored then ${\cal
O}(1/{m}_{Q}^{2})$ corrections to  semileptonic decay rates at zero
recoil are small.  But we are led to a  very different result when
hyperfine effects are included.

Our model Lagrangian will contain terms which couple the heavy meson
fields to the heavy and light quark fields. \begin{equation}
-\overline{Q(x)}\left\{i\gamma_{5}P(x)F_{P}(x-
y)+\gamma_{\mu}V^{\mu}(x)F_{V}(x-y)\right\}q(y)+{\rm h.c.} \label{01}
\end{equation} $Q$ and $q$ are the heavy and light quarks contained
in the heavy  pseudoscalar and vector mesons $P$ and $V$.
${F}_{P,V}(x-y)$ are  damping factors which will suppress large
momentum flow into the light  quark.  This is the essential physical
effect of the light quark wave  function.  In momentum space we take
\begin{equation}
F_{P,V}(k)=\frac{Z_{P,V}^{2}}{-k^{2}+\Lambda_{P,V}^{2}} \label{03}
\end{equation} where $k$ is the momentum of the light quark.  The
quantities ${Z}_{P,V}$ and  ${\Lambda }_{P,V}$ are completely
determined for each meson in a manner  to be described below.   In
particular the fact that ${\Lambda }_{P}$ is not equal to  ${\Lambda
}_{V}$ reflects the spin dependent distortion of ``wave  functions"
mentioned above, and ${\Lambda }_{P,V}$ will be fixed by the
physical,  spin dependent meson masses.  Such a connection between
the distortion and the meson  mass splitting will be present in any
quark model of heavy mesons, whether or not the  model explicitly
relates hyperfine effects to gluon exchange.

In \cite{2} we considered the effect of raising $F_{P,V}$ in
(\ref{03})  to powers $n$ other than unity.  We found that meson
decay constants were the  physical quantities most sensitive to the
high energy behavior of the vertex factors,  and we found that they
decrease for increasing $n$.  To obtain realistic meson decay
constants, values of $n$ close to but greater than 1 are preferred.
Although $n=1$  yields a divergent vector meson decay constant, the
$n=1$ case modified by inverse  powers of logarithms would probably
also be acceptable.  Such modifications would  have minimal impact on
the quantities of interest in this paper, and we shall focus on  the
pure $n=1$ case as given by (\ref{03}).

The rest of our model Lagrangian consists of standard kinetic and
mass  terms for the quarks.  The heavy mesons are composed of quarks,
and thus  to avoid double counting of dynamical degrees of freedom
the model  Lagrangian does not contain meson kinetic terms.  Neither
does it contain  mass terms or meson self-couplings; all these terms
are to be modeled  via the quark loop.  Various other fields may be
coupled to the quarks,  depending on the application.  For example to
describe semileptonic decay  we will couple the heavy quarks to the
appropriate external gauge fields.   Processes involving pions or
kaons may be described by coupling these  fields to the light quarks.

The quark masses are the only parameters of the model.  For the heavy
quarks we use the values $m_{b}=4.8$ GeV and $m_{c}=1.44$ GeV.  For
the  light quark we use its constituent mass ${m}_{q}$; but here
there is some  uncertainty.  Since we expect that the actual momentum
dependent  constituent quark mass has fallen somewhat at momenta
typical in the loop, we take a  (momentum independent)
${m}_{q}\approx 250$ MeV.  We will also explore the  sensitivity of
results to changes in ${m}_{q}$.

The free quarks may be integrated out of the theory and this yields
an  effective theory of heavy mesons.  The form of such a theory has
been well  described in the literature.\cite{3}  This is a theory of
heavy mesons  propagating on mass shell or close to mass shell.  It
does not describe  diagrams with closed loops of heavy mesons.  It
may describe pion loop  corrections with an internal heavy meson line
since the effective cutoff  on such a loop is far below the heavy
meson mass.

Here we note a peculiar feature of our model.  It turns out that upon
carrying out the procedure just described, the resulting effective
heavy  meson Lagrangian has an overall nonstandard minus sign.  But
this sign,  which would be disastrous for a normal quantum field
theory, turns out to  be an unphysical sign for this effective
theory.  This is due to the absence  of heavy meson loops.  With each
heavy meson vertex and each heavy  meson propagator having an
additional minus sign, the result is that a  diagram receives a sign
${(-1)}^{j}$ where $j$ is the number of heavy  meson lines running
through the diagram.  Such a sign is of no consequence for  physical
quantities calculated from these amplitudes.\footnote{One may wonder
about  the existence of negative energies.  But terms of higher order
in derivatives and heavy  quark mass would be relevant for the
Hamiltonian.}  The amplitudes of interest here  are determined by the
tree level two-meson-external-gauge-field vertices with heavy  mesons
on shell.  In  the following we will simply omit the minus sign to
coincide  with standard conventions.

We denote the proper pseudoscalar and vector self-energy graphs by
$i\Gamma_{P}(p^{2})$ and $-ig_{\mu\nu}\Gamma_{V}(p^{2})+\ldots$,
where the ellipsis denotes $p_{\mu}p_{\nu}$ terms.  Because the model
treats the quarks as free, the $\Gamma_{P,V}$ become complex above
the  threshold at $p^{2}=(m_{Q}+m_{q})^{2}$.  We will ignore this
consequence  of free quarks and drop all such imaginary pieces of
quark loop graphs.\footnote{The imaginary parts may in fact be kept
for all quantities we calculate; the result would be a consistent
description of mesons coupling to free quarks.  Our dropping the
imaginary parts in the end is an effort to incorporate confinement.
In this connection we note that the confinement physics of QCD occurs
at a lower momentum scale than the scale characterizing our model,
$\Lambda_{P,V} \approx 650-700$ MeV.}   The meson masses are defined
by $\Gamma_{P,V}(m_{P,V}^{2})=0$.  The  mass functions ${\Gamma
}_{B,{B}^{{}^*}}$ are plotted versus $\sqrt{p^{2}}$ in  Fig. 1.  The
zeros lie between the threshold kink and  singularities at
$p^{2}=(m_{Q}+\Lambda_{P,V})^{2}$.  (We therefore obtain  sensible
meson masses only if $\Lambda_{P,V}>m_{q}$.)  The  $\Lambda_{P,V}$
are determined by requiring that the zeros of the meson  mass
functions coincide with the experimental meson masses.

The constants $Z_{P,V}$ are determined by normalizing in either of
two  equivalent ways.  The Ward identities relate the slopes of the
respective  mass functions at their zeros to the $q^{2}=0$ matrix
elements of the  heavy quark vector current between on-shell mesons.
Normalizing these matrix  elements to unity is equivalent to the
condition  $\Gamma_{P,V}^{\prime}(m_{P,V}^{2})=1$, where the prime
denotes  differentiation with respect to $p^{2}$.  Thus,
\begin{equation} \Gamma_{P,V}(p^{2})\approx p^{2}-m_{P,V}^{2}
\label{04} \end{equation} in the neighborhood of $p^{2}=m_{P,V}^{2}$,
corresponding to standard  meson kinetic terms.  For the $B$ and $D$
sectors, the values of the input  meson masses and the resulting
values of $\Lambda_{P,V}$ and $Z_{P,V}$  are shown in the top half of
Table \ref{table1}.

\begin{table}  \begin{center} \begin{tabular}{c|ccc||c|ccc} $P$ &
$m_{P}$ & $\Lambda_{P}$ & $Z_{P}$ &  $V$ & $m_{V}$ & $\Lambda_{V}$ &
$Z_{V}$ \\ \hline $B$ & 5.279 & 0.6214 & 1.1026 & $B^{\ast}$ & 5.325
& 0.7015 & 1.2896 \\ $D$ & 1.869 & 0.5367 & 0.7446 &  $D^{\ast}$ &
2.010 & 0.7623 & 1.1927 \\ \hline $B$ & 5.31 & 0.6661 & 1.1920 &
$B^{\ast}$ & 5.31 & 0.6789 & 1.2432 \\ $D$ & 1.97 & 0.6739 & 0.9808 &
$D^{\ast}$ & 1.97 & 0.7045 & 1.0878 \end{tabular} \end{center}
\caption{Input meson masses and resulting values of $\Lambda_{P,V}$
and  $Z_{P,V}$ with (top half of table) and without (bottom half)
hyperfine splitting (in GeV).\label{table1}} \end{table}

We are interested in the effects associated with hyperfine mass
splitting,  $m_{V}-m_{P}$.  In the model this originates largely in
the different  values of ${\Lambda }_{V}$ and ${\Lambda }_{P}$.
(There is also a little  mass splitting even in the case ${\Lambda
}_{V}={\Lambda }_{P}$, with the  vector mass slightly {\it less} than
the pseudoscalar mass.)  We will be  comparing our results based on a
fit to physical meson masses with  results in the case of no
hyperfine splitting.  We will take the latter case  to correspond to
the values of ${\Lambda }_{P}$ and ${\Lambda }_{V}$  which yield both
the pseudoscalar and vector masses equal to
${\frac{1}{4}}{m}_{P}+{\frac{3}{4}}{m}_{V}$.  The amounts by which
the  physical masses are shifted from this common mass is
characteristic of  the effect of hyperfine mass splitting.  We
display for comparison  purposes the various input parameters for the
case of no hyperfine  splitting in the bottom half of Table
\ref{table1}.

We are also interested in comparing to results obtained at first
order in  the heavy quark expansion.\cite{2}  The heavy quark limit
of the model is  described by the conditions \begin{equation}
\Lambda_{P,V}\rightarrow \Lambda \;,\; \Lambda/m_{Q}\rightarrow 0
\;,\; \Lambda/m_{q} \; {\rm fixed}. \end{equation} This common
$\Lambda$ is the same for all heavy quark flavors.  At zeroth  order
in an expansion in $\Lambda/m_{Q}$, quantities depend only on the
ratio $\Lambda/m_{q}$ and are independent of the way in which
$\Lambda_{P,V}\rightarrow \Lambda$.  At first order in the expansion
we  write \begin{equation} {\Lambda }_{P,V}=\Lambda
\left({1-(\delta_{P,V} h+ g){\frac{\Lambda  }{{m}_{Q}}}}\right)
\end{equation} where ${\delta}_{P}=3$ and ${\delta}_{V}=-1$. In
\cite{2} we study the dependence of first order corrections on the
parameters $\Lambda/m_{q}$, $g$ and $h$.\footnote{Our definition of
$g$ and $h$ here differs from that in \cite{2}.}

The parameter $h$ is responsible for most of the hyperfine mass
splitting.  We may choose a set of parameters in the first order
model  which produces an optimal meson mass spectrum.  For
$m_{q}=250$ MeV  such a set is $\Lambda =667$ MeV, $g=-0.13$ and
$h=0.19$ and it yields  $(B,{B}^{{}^*})$ and $(D,{D}^{{}^*})$ masses
within 0.2\% and 2\%  respectively of the physical masses.

The form factors for the semileptonic decays $B\rightarrow D$ and
$B\rightarrow D^{\ast}$ are defined as follows. \begin{equation}
\left\langle{{P}_{2}({v}_{2})\left|{{V}_{\mu
}}\right|{P}_{1}({v}_{1})}\right\rangle=\sqrt
{{M}_{{P}_{1}}{M}_{{P}_{2}}}\left[{{h}_{+}(\omega
){({v}_{1}+{v}_{2})}_{\mu }+{h}_{-}(\omega ){({v}_{1}-{v}_{2})}_{\mu
}}\right] \label{05} \end{equation} \begin{equation}
\left\langle{{V}_{2}({v}_{2})\left|{{V}_{\mu
}}\right|{P}_{1}({v}_{1})}\right\rangle=\sqrt
{{M}_{{P}_{1}}{M}_{{V}_{2}}}{h}_{V}(\omega ){\varepsilon }_{\mu  \nu
\rho \sigma }{\varepsilon }_{2}^{\ast \nu }{v}_{2}^{\rho
}{v}_{1}^{\sigma } \label{06} \end{equation} \begin{eqnarray}
\lefteqn{\left\langle{{V}_{2}({v}_{2})\left|{{A}_{\mu
}}\right|{P}_{1}({v}_{1})}\right\rangle=} \\ \nonumber &&-i \sqrt
{{M}_{{P}_{1}}{M}_{{V}_{2}}}\left[{(\omega  +1){h}_{{A}_{1}}(\omega
){\varepsilon }_{2\mu }^{\ast }-\left({{h}_{{A}_{2}}(\omega
){v}_{1\mu  }+{h}_{{A}_{3}}(\omega ){v}_{2\mu }}\right){\varepsilon
}_{2}^{\ast }\cdot  {v}_{1}}\right] \label{07} \end{eqnarray} The
$v$'s are meson velocities, and $\omega ={v}_{1}\cdot {v}_{2}$.

At zeroth order in the heavy quark expansion, \begin{equation}
{h}_{+}(\omega )={h}_{V}(\omega )={h}_{{A}_{1}}(\omega
)={h}_{{A}_{3}}(\omega )=\xi (\omega )\;,\; {h}_{-}(\omega
)={h}_{{A}_{2}}(\omega )=0. \label{08} \end{equation} $\xi (\omega)$
is the Isgur-Wise function.  The first order corrections to the
$h_{i}(\omega)$ require the introduction of only four additional
universal  functions.\cite{4}  In \cite{2} we show that this is also
true for our model, for any  choice of parameters.  This is a
nontrivial test for any model of heavy mesons.

We may calculate the $h_{i}(\omega)$ in the full model from the
appropriate three point quark loop graphs.  Our results are displayed
in Fig.  2.  For comparison we also display $\xi (\omega  )$ emerging
at zeroth order in the model (using $\Lambda =667$  MeV).  We find
that ${h}_{V,A_{3},A_{1},+}(\omega)$ display  substantial shifts of
varying amounts from $\xi (\omega )$, although the  shape of these
functions is fairly similar.  In addition ${h}_{- ,A_{2}}(\omega )$
deviate significantly from zero.

We collect the full zero recoil values ${h}_{i}(1)$ in the first row
of Table  \ref{table2}, and in the second row we collect the zero
recoil values with  no hyperfine splitting, $h_{i}^{{\rm
no\,hyp}}(1)$.  Most of the $h_{i}(1)$  deviate significantly further
from their values in the heavy quark limit  than do the
${h}_{i}^{{\rm no\,hyp}}(1)$.  Especially interesting are the
quantities ${h}_{+}(1)$ and ${h}_{{A}_{1}}(1)$ which are both still
equal to  unity in the heavy quark expansion at first order (Luke's
Theorem \cite{4}).  We see  from Table \ref{table2} that at higher
orders, in the absence of hyperfine splitting,  ${h}_{+}(1)$ and
${h}_{{A}_{1}}(1)$ remain very close to unity.  But the higher  order
corrections {\it with} hyperfine splitting cause significant
deviation in  these quantities away from the heavy quark limit.

\begin{table}  \begin{center} \begin{tabular}{c|cccccc} $i$ & $V$ &
$A_{3}$ & $A_{1}$ & $+$ & $-$ & $A_{2}$ \\ \hline $h_{i}(1)$ & 1.401
& 1.307 & 1.155 & 1.107 & $-$.138 & $-$.249 \\ $h_{i}^{{\rm
no\,hyp}}(1)$ & 1.232 & 1.149 & 1.001 & 0.994  & $-$.108 & $-$.274 \\
$h_{i}^{{\rm 1st}}(1)$ & 1.228 & 1.152 & 1 & 1 & $-$.124 & $-$.270 \\
\hline $h_{i}^{\prime}(1)$ & $-$2.21 & $-$1.95 & $-$1.71 & $-$1.83 &
0.20 & 0.51  \\ $(h_{i}^{{\rm no\,hyp}})^{\prime}(1)$ & $-$1.76 &
$-$1.54 & $-$1.32 & $- $1.53  & 0.14 & 0.51 \\ $(h_{i}^{{\rm
1st}})^{\prime}(1)$ & $-$1.69 & $-$1.49 & $-$1.29 & $-$1.54  & 0.13
&  0.46 \end{tabular} \end{center}  \caption{Zero recoil values of
form factors $h_{i}$ (top half of table) and derivatives
$h_{i}^{\prime}$ (bottom half) with hyperfine splitting, without
hyperfine splitting, and at first order.\label{table2}} \end{table}

The zero recoil values from the first order model, $h_{i}^{{\rm
1st}}(1)$,  are given in the third row of Table \ref{table2}.  In
\cite{2} we found that  these values are independent of $g$ and $h$
and thus independent of  hyperfine splitting.  The substantial
difference between ${h}_{i}(1)$ and  ${h}_{i}^{{\rm no\,hyp}}(1)$ is
therefore an ${\cal O}(1/{m}_{Q}^{2})$  effect.  It is also
interesting that the $h_{i}^{{\rm 1st}}(1)$ are very close  to the
${h}_{i}^{{\rm no\,hyp}}(1)$.  This shows that were it not for the
hyperfine effects, the corrections beyond first order would be very
small.   It appears that hyperfine effects completely dominate the
${\cal  O}(1/{m}_{Q}^{2})$ corrections.

In the bottom half of Table \ref{table2} we give the values of the
slopes  of the form factors at zero recoil.  By comparing the first
two rows we  see that the effect of hyperfine splitting is to make
the slopes more  negative.  And the effect of hyperfine splitting at
first order (that is, the  shift in the $(h_{i}^{{\rm
1st}})^{\prime}(1)$ caused by nonzero  $h$) is no more than about
1\%.  Thus again the main effect of hyperfine  splitting occurs at
${\cal O}(1/{m}_{Q}^{2})$.

In Table \ref{table3} we show how the $h_{i}(1)$ change when the
light  quark mass ${m}_{q}$ is varied.  Increasing it to 300 MeV
increases the  corrections even further.  And the corrections remain
substantial even if  ${m}_{q}$ is lowered arbitrarily. \begin{table}
\begin{center} \begin{tabular}{c|cccccc} $m_{q}({\rm MeV})$ & $V$ &
$A_{3}$ & $A_{1}$ & $+$ & $-$ & $A_{2}$  \\  \hline 300 & 1.481 &
1.390 & 1.225 & 1.156 & $-$.150 & $-$.259 \\ 200 & 1.363 & 1.263 &
1.123 & 1.086 & $-$.132 & $-$.242 \\ 10 & 1.334 & 1.193 & 1.103 &
1.082 & $-$.124 & $-$.208 \end{tabular} \end{center}
\caption{Dependence of $h_{i}(1)$ on light quark mass
$m_{q}$.\label{table3}}\end{table}

We consider further the two quantities which are protected from first
order  corrections at zero recoil.  The following relations are
derived in \cite{6}. \begin{equation}
{h}_{+}(1)=1+{\left({{\frac{1}{2{m}_{c}}}-
{\frac{1}{2{m}_{b}}}}\right)}^{2}{\ell  }_{1}(1)+{\cal
O}\left({{\frac{1}{{m}_{Q}^{3}}}}\right) \label{21} \end{equation}
\begin{equation} {h}_{{A}_{1}}(1)=1+\left({{\frac{1}{2{m}_{c}}}-
{\frac{1}{2{m}_{b}}}}\right)\left[{{\frac{1}{2{m}_{c}}}{\ell }_{2}(1)-
{\frac{1}{2{m}_{b}}}{\ell }_{1}(1)}\right]+{\frac{\Delta
}{4{m}_{c}{m}_{b}}}+{\cal O}\left({{\frac{1}{{m}_{Q}^{3}}}}\right)
\label{22} \end{equation} In a nonrelativistic quark model ${\ell
}_{1}(1)$ and ${\ell }_{2}(1)$ are  negative since they may be
related to the deficit in a wave function overlap
integral.\footnote{We find from \cite{7} ${\ell }_{1}(1) = {\ell
}_{2}(1) \approx -{\frac{3}{4}}{m}_{q}^{2}$ , i.e. $1/4$ times those
in \cite{6}.}  This is not the case for our relativistic quark
model.  Our meson-$q$-$\overline{Q}$ vertex factors are analogous to
wave functions, but they  are inserted into a relativistic quark loop
calculation.  Our results explicitly show that  as the two ``wave
functions" are distorted by hyperfine effects and made different
from each other, the result is a positive correction.  We have
numerically isolated the  $1/{m}_{Q}^{2}$ corrections in
(\ref{21},\ref{22}) and find $\Delta \approx (1.6\,{\rm GeV})^{2}$,
${\ell }_{1}(1)\approx (1.0\,{\rm GeV})^{2}$, and
${\ell}_{2}(1)\approx (0.6\,{\rm GeV})^{2}$.  The total magnitude of
the $1/{m}_{Q}^{2}$ corrections in (\ref{21},\ref{22}) amounts to
about $2/3$ of the corrections to all orders.

The $\Delta $ term in (\ref{22}) may be written as $1.5{\Lambda
}^{2}/{m}_{b}{m}_{c}$; the dimensionless coefficient here does not
appear to be unreasonably large.  In \cite{6} $\Delta $ is written as
\begin{equation} \Delta ={\frac{4}{3}}{\lambda }_{1}+2{\lambda
}_{2}+\cdots \end{equation}  where the ellipsis represents
corrections resulting from a double insertion of the chromo-magnetic
operator.  The $\lambda$'s are defined by \begin{equation} {\lambda
}_{1}=-{\frac{1}{4}}(\delta {m}_{P}^{2}+3\delta
{m}_{V}^{2})~~~\mbox{and}~~~{\lambda }_{2}=-{\frac{1}{4}}(\delta
{m}_{P}^{2}-\delta {m}_{V}^{2}) \end{equation} where \begin{equation}
{m}_{M}-{m}_{Q}=\overline{\Lambda }+{\frac{\delta
{m}_{M}^{2}}{2{m}_{Q}}}+\cdots \end{equation}  We find in our model
that $\overline{\Lambda }=0.50$ GeV, ${\lambda }_{1}=-(0.27\,{\rm
GeV})^{2}$ and ${\lambda }_{2}=(0.29\,{\rm GeV})^{2}$.  These
$\lambda$'s are small contributions which largely cancel in $\Delta$,
and thus the bulk of contribution to $\Delta$ must originate in the
double insertions of the chromo-magnetic operator.  In \cite{6} it was
assumed that these contributions could be neglected.

We now turn to the two processes which, by observing at zero recoil,
have  been proposed \cite{5} as fairly model independent methods for
extracting  ${V}_{cb}$.  In the differential decay rate at zero
recoil for the process  $B\rightarrow Dl \nu $ the following factor
appears \begin{equation}
{\left|{{h}_{+}(1)-{\frac{{m}_{B}-{m}_{D}}{{m}_{B}+{m}_{D}}}{h}_{-
}(1)}\right|}^{2}{}^{} \label{09} \end{equation} The deviation of
this quantity from unity gives the correction to the heavy  quark
limit.  This might be expected to be small since the only first order
$1/{m}_{Q}$ correction occurs in ${h}_{-}(1)$ and this is suppressed
by  the prefactor.  But inserting values from Table \ref{table2}
gives a  positive 38\% correction with hyperfine splitting and a 9\%
correction  without hyperfine splitting.  For the decay $B\rightarrow
D^{{}^*}l\nu $ the  factor of interest is
${\left|{{h}_{{A}_{1}}(1)}\right|}^{2}$ which has no  first order
$1/{m}_{Q}$ corrections.  Here the correction is 33\% with  hyperfine
splitting, to be compared with essentially no correction for no
hyperfine splitting.

These numerical values for the corrections should be considered as
illustrative.  They  depend not only on the light constituent mass,
but also on the $c$ and $b$ quark  masses we have chosen.  And the
corrections obviously reflect the simple form we  have chosen for the
meson vertex factors in (\ref{03}).  But the existence of
significant corrections from hyperfine effects, with signs as given,
seems to be generic  to this class of relativistic quark models.

In this paper we have argued by way of an explicit model that
corrections  to the heavy quark limit may be larger than previously
thought.  This is  seen most graphically in Fig. 2 which depicts the
form factors for  semileptonic decay of heavy mesons.  At zero recoil
the larger than  expected corrections appear at ${\cal
O}(1/{m}_{Q}^{2})$ in the heavy  quark expansion.  We have traced the
origin of these corrections to the  physics associated with hyperfine
mass splitting.  We hope that these  results will encourage further
study of hyperfine effects in  nonrelativistic quark models and sum
rule calculations.

\vspace{2ex} \noindent {\bf Acknowledgment} \vspace{1ex}

We thank M. Wise, P. O'Donnell, J. Moffat, and T. Morozumi for useful
discussions.  This research was supported in part by the Natural
Sciences  and Engineering Research Council of Canada.

\newpage \noindent{\bf Figure Captions} \vspace{6ex}

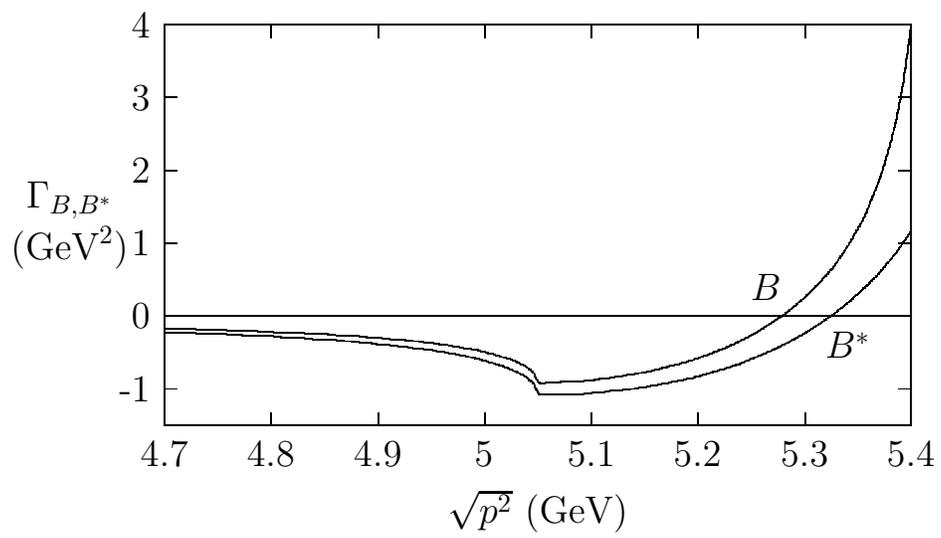
\noindent{Figure 1. Normalized $B$ and $B^{\ast}$ mass functions for
$m_{q}=250$ MeV.} \vspace{3ex}

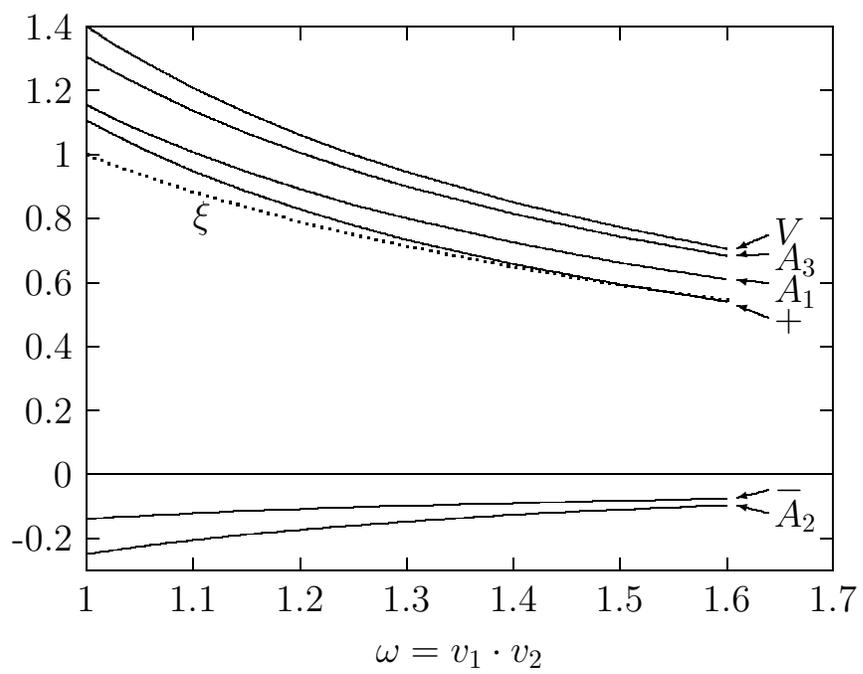
\noindent{Figure 2. $B\rightarrow D$ and $B\rightarrow D^{\ast}$ form
factors $h_{i}(\omega)$ and Isgur-Wise function $\xi(\omega)$.}
\vspace{3ex}

\end{document}